# Patterned and Functionalized Nanofiber Scaffolds in 3-Dimensional Hydrogel Constructs Enhance Neurite Outgrowth and Directional Control

Richard J. McMurtrey[1,2]

[1] Institute of Biomedical Engineering, Department of Engineering Science, Old Road Campus Research Building, University of Oxford, Oxford, OX3 7DQ, United Kingdom Email: richard.mcmurtrey@eng.oxon.org

[2] Institute of Neural Regeneration & Tissue Engineering, Highland, UT 84003, United States, Email: richard.mcmurtrey@neuralregeneration.org

## Abstract

*Objective*: Neural tissue engineering holds incredible potential to restore functional capabilities to damaged neural tissue. It was hypothesized that patterned and functionalized nanofiber scaffolds could control neurite direction and enhance neurite outgrowth.

*Approach*: A method of creating aligned electrospun nanofibers was implemented and fiber characteristics were analyzed using environmental scanning electron microscopy. Nanofibers were composed of polycaprolactone (PCL) polymer, PCL mixed with gelatin, or PCL with a laminin coating. 3-dimensional hydrogels were then integrated with embedded aligned nanofibers to support neuronal cell cultures. Microscopic images were captured at high-resolution in single and multi-focal planes with eGFP-expressing neuronal SH-SY5Y cells in a fluorescent channel and nanofiber scaffolding in another channel. Neuronal morphology and neurite tracking of nanofibers were then analyzed in detail.

*Main Results*: Aligned nanofibers were shown to enable significant control over the direction of neurite outgrowth in both 2-dimensional (2D) and 3-dimensional (3D) neuronal cultures. Laminin-functionalized nanofibers in 3D hyaluronic acid (HA) hydrogels enabled significant alignment of neurites with nanofibers, enabled significant neurite tracking of nanofibers, and significantly increased the distance over which neurites could extend. Specifically, the average length of neurites per cell in 3D HA constructs with laminin-functionalized nanofibers increased by 66% compared to the same laminin fibers on 2D laminin surfaces, increased by 59% compared to 2D laminin-coated surface without fibers, and increased by 1052% compared to HA constructs without fibers. Laminin functionalization of fibers also doubled average neurite length over plain PCL fibers in the same 3D HA constructs. In addition, neurites also demonstrated tracking directly along the fibers, with 66% of neurite lengths directly tracking laminin-coated fibers in 3D HA constructs, which was a 65% relative increase in neurite tracking compared to plain PCL fibers in the same 3D HA constructs and a 213% relative increase over laminin-coated fibers on 2D laminin-coated surfaces.

*Significance*: This work demonstrates the ability to create unique 3-dimensional neural tissue constructs using a combined system of hydrogel and nanofiber scaffolding. Importantly, patterned and biofunctionalized nanofiber scaffolds that can control direction and increase length of neurite outgrowth in 3-dimensions hold much potential for neural tissue engineering. This approach offers advancements in the development of implantable neural tissue constructs that enable control of neural development and reproduction of neuroanatomical pathways, with the ultimate goal being the achievement of functional neural regeneration.

## Introduction

Damage to neural tissue is one of the leading causes of death and permanent disability in the world and also presents one of the greatest challenges in current medical care. The burdens of neural tissue dysfunction are carried by millions of people who suffer conditions of stroke, traumatic brain injury, spinal cord injury, nerve injury, and neurodegenerative diseases like Parkinson's disease, Alzheimer's disease, Huntington's disease, and amyotrophic lateral sclerosis. Neural tissue engineering holds potential to restore functional capabilities to damaged neural tissue, thereby offering the hope of greatly improving quality of life, but the design of neural tissue constructs presents many challenges. The regeneration and restoration of neural tissue requires many important cellular and extracellular components, which interact with each other in several important ways. Extracellular materials serve as scaffolding for cellular architecture and can provide many biochemical signals that influence stem cell differentiation and cell behavior, and cells themselves also interact with each other to facilitate neural function, provide trophic support, and direct differentiation during development [1]. Guiding cells to achieve the intended goal of survival, proliferation, differentiation, and network formation is a difficult challenge, but integration of multiple components will provide the best approach to engineering functional neural tissue.

For neuronal cells, three-dimensional (3D) culture is essential to reconstruct the innate structure-function relationship of neuronal tissue. Furthermore, 3D constructs enable structured transplantation into an *in vivo* environment of damaged neural tissue, and 3D cultures also more realistically reconstruct cellular interactions and adherence with surrounding cells and matrix and can also replicate the mass transfer characteristics of neuronal tissue, which may be important for preparing cells for survival after implantation [2]. Hydrogel matrices provide a useful method for creating 3D cultures, and they have many properties that make them useful for neural tissue engineering, including a low elastic modulus with stiffness adjustable to that of neural tissue and an array of possible compositions with various biocompatible polymers [3-10]. Hydrogels may also aid the ability of neurons to extend new axons [10-16]. Hyaluronic acid (HA) in particular is a biocompatible and biodegradable polymer naturally found in neural tissue, and hyaluronan hydrogels are tunable in stiffness and degradation properties through polymer density and crosslinking [3,17-18]. In addition, hyaluronic acid hydrogels have been shown to support seeded cells and provide certain physical and biological cues that influence neurite extension and differentiation of neural precursors [19-23].

Nanofibers can be created using electrospinning, a technique used to produce fibers that can range in diameter from nanometers to micrometers [24-27]. Electrospun fibers can be made of many types of polymers and composite combinations that can be utilized to imitate properties of native extracellular matrix or to optimize various chemical, mechanical, electrical, architectural, and biological properties [5-6,25-39]. Polycaprolactone (PCL), for example, is a flexible, biocompatible, and biodegradable polymer that has been investigated for use in many biological applications [6,27,36,40]. Electrospun fibers have been used to create scaffolds for cell culture, and experiments have shown that unique effects can be achieved by culturing neuronal cells on various micropatterned scaffolds [27,33,41-47]. In particular, studies have shown that electrospun fiber scaffolding can exert significant effects on neuronal development [28-29,35,45,48-51] and neurite formation [25,30-31,52-55]. This includes evidence that neuronal cells exhibit neurite outgrowth highly oriented to underlying fibers [25-26,30,32,34,38,48,56-62] and that peripheral neurons may extend neurite length by 20% when aligned along electrospun fiber surfaces [30]. Biological molecules like laminin, collagen, and others have also been shown to enhance adhesion and tracking of neurites along underlying polymer substrates [6,34-35,37,57-58,63]. Much of the work on nanofiber scaffolds, however, has only examined neural growth on dense fiber collections on two-dimensional surfaces.

It was hypothesized that the synthesis of aligned electrospun fibers within 3-dimensional hydrogels could potentially create several advantageous characteristics for neural tissue constructs, such as controlled direction of neurite extension, scaffolding that could serve as an attachment point for cells in a low-density hydrogel, synthesis of mechanical and biological signals into a single construct, and a support architecture for replicating neuroanatomical pathways. This approach was therefore investigated as a potential avenue for restructuring the neuroanatomical pathways and patterned neural circuitry that exist in the brain and spinal cord. This work first explored optimization of nanofiber manufacture, and then used these fibers to produce an aligned plane of parallel electrospun fibers within a 3-dimensional hydrogel construct, with the hypothesis that such scaffolds could be employed to support neuronal cultures and guide neurite development. More specifically, it was hypothesized that nanofiber constructs could provide topographical guidance for neuronal cell attachment and neurite growth in a 3-dimensional environment, and it was also explored whether functionalization of nanofibers with gelatin or laminin could enhance features of neural development such as neurite length and direction.

<p style="text-align:center"><strong>Methods and Materials</strong></p>

*Electrospinning Fibers*

Electrospinning parameters were set up to produce straight homogeneous fibers in parallel alignment, as described in the appendix. Fibers were optimized for strength and robustness in 3-dimensional wells of culture media, with polymer composition of 6% (w/v) poly-caprolactone (PCL) in dichloromethane (Sigma-Aldrich), with a PCL molecular weight of 70-90 kD (Sigma-Aldrich). Fibers were collected on a grounded drum rotating at approximately 1000 rpm at a 10 cm distance from the microinfuser with a 10 kV potential difference.

In addition to fibers made of PCL alone, 6% (w/v) PCL was also mixed with gelatin (denatured porcine collagen) as a 25% w/w gelatin to PCL polymer composite, making the total polymer concentration 7.5% (w/v) in the electrospinning solution. Finally, another group of PCL fibers was made by coating the PCL nanofibers with laminin (Sigma-Aldrich) by placing the electrospun fibers in a liquid solution of laminin (10 μg/ml) for 2 hours then washing twice with PBS.

Electrospun fibers were characterized using environmental scanning electron microscopy (Carl Zeiss Evo LS15 VP SEM). Fiber samples were desiccated and loaded onto carbon samples, and photographs were obtained at 100-20,000X magnification with appropriate scale bars. The mean width and standard deviation are reported, and fibers widths were compared using two-tailed t-tests assuming unequal variances.

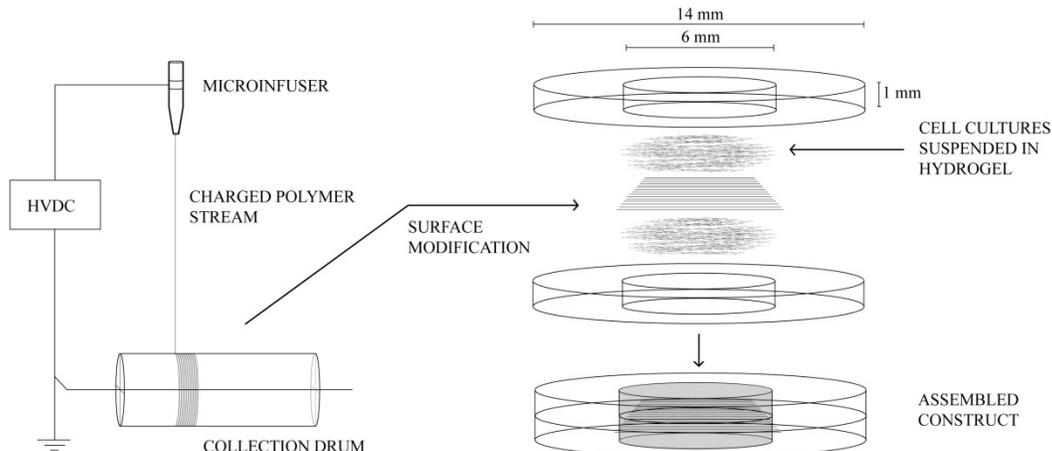

Figure 1: Schematic of electrospinning fibers and creation of the 3-dimensional construct. The plane of aligned nanofibers was sandwiched between two PDMS rings, and the PDMS wells were filled with cellularized hydrogel.

*Scaffolding Construction*

Culture wells were created as PDMS rings, as pictured in Figure 1. A two part kit (Dow Corning Sylgard 184) was used to create flat PDMS disks, using a 10:1 w/w ratio of PDMS to curing agent. The mixture was cured at 60°C for 4 hours. PDMS rings were fashioned to be 1 mm in height, 14 mm outer diameter, with a 6 mm diameter inner well. A layer of aligned fibers was spread over the top of the first ring, spanning the inner diameter space, and this ring was placed on top of an autoclaved glass coverslip base, after which another ring was stacked on top of the first ring, with fibers attached between the junction of the two rings. The ring assemblies were placed in 24-well plates, and the constructs were disinfected with 15 minutes of UV-C exposure, which did not grossly affect the fibers, although this process is known to cause polymer chain fragmentation in PCL and other polymers, which may result in weakening of the fibers in a dose-dependent fashion [64].

In addition, previous experiments have cultured neurons on bundled or dense networks of 2-dimensional aligned nanofibers, but such dense networks in 3-dimensions can obscure imaging and make cell-fiber interactions ambiguous, and dense fiber bundles can also force the neuronal cells to settle onto and migrate along fibers without necessarily having a particular affinity for them. In order to overcome these issues, nanofibers were spaced more sparsely, approximately 5-20 microns apart such that cells could generally touch and pass through the fibers. This allowed careful interpretation of whether a neurite was directly tracking a fiber, allowed cells to either be captured on the fibers or pass through the fibers, and allowed adequate lighting and imaging through the fiber layer.

*Cell Culture*

In order to produce a neuronal line without glial cells to affect growth or imaging in 3D culture, neuronal cells were induced from an SH-SY5Y cell line transfected with constitutively active enhanced green fluorescent protein (eGFP). Similar to previous SH-SY5Y protocols [65-66], the cells were cultured in complete media (RPMI 1640 media with sodium bicarbonate and L-glutamine (Sigma) supplemented with 10% FBS, 1% Minimum Essential Medium (MEM), 50 U/ml penicillin, and 50 μg/ml streptomycin) at 37°C with 5% $CO_2$, with feedings every 2-3 days, until cells were 75-80% confluent. To differentiate into functionally mature cells of neural phenotype, the media was replaced with complete medium supplemented with 10μM all-trans-retinoic acid, BDNF (50 ng/ml), NGF (10 ng/ml), & 2% B27 supplement for 5 days prior to seeding in the 3D constructs, and non-adherent cells were washed away in this process. Cells were kept at 37°C with 5% $CO_2$ and were fed again at the half-way point of 5 days. At the end of 5 days, cells were washed twice in PBS, then detached from the flask by adding 5ml of trypsin for 5 minutes at 37°C, after which cells were washed from the flask surface and mixed with 5ml complete medium to produce a 10ml suspension of cells. The cell suspension was centrifuged at 300g for 4 minutes to form a pellet, the supernatant removed, and the cells were re-suspended in 10ml of complete medium. A cell count was performed using a hemocytometer and aliquots were made for each condition according to the numbers described below. Neuronal identity was confirmed with immunostaining for neuron-specific class III beta-tubulin using TUJ-1 monoclonal antibody (R&D Systems) at 1:1000 dilution and Alexa-594 goat anti-mouse secondary antibody (Life Technologies) at 1:500 dilution. After fixating cells with 4% paraformaldehyde and washing with PBS, cells were permeabilized and blocked with 0.2% Triton X-100 and 5% BSA (Sigma-Aldrich) in PBS for 1.5 hours, and primary antibodies were applied for 1.5 hours in a wash buffer of 0.2% BSA and 0.2% Triton X-100 in PBS. Cells were then washed with wash buffer and blocked with block buffer for 30 minutes at room temperature, and secondary antibodies were applied for 1.5 hours and then washed with wash buffer.

*Integration of Cells and Constructs*

A hierarchical set of control conditions was used, which included 2D laminin-coated glass coverslips (for comparing conditions of cells seeded on 2D surfaces versus 3D hydrogels), 3D hydrogels without fibers (for comparing to conditions of 3D hydrogels with fibers), and 3D hydrogels with uncoated electrospun fibers (for comparing conditions of 3D hydrogels with laminin-functionalized fibers). Fibers consisted of either plain PCL polymer alone, PCL polymer with gelatin composite, or PCL polymer fibers coated with laminin, as described previously. All conditions were assembled in a cell culture hood using sterile technique.

For 2D conditions, glass coverslips were coated with laminin by exposing them to a 10 μg/ml laminin solution for 2 hours and washing with PBS. A set of glass coverslips was also left uncoated as another control condition to compare untreated surface attachment of cells, but cells did not adhere or grow on this surface. For 3D conditions, two distinct types of hydrogel were studied: hyaluronic acid (HA), which is relatively inert and has minimal cell adhesion properties, and Engelbreth-Holm-Swarm (EHS) extracellular matrix, which originates from the secretions of a murine sarcoma line and is rich in cell attachment points. The HA hydrogels were made from HyStem kits (Glycosan), with the composition diluted to 0.75% from the standard 1% using the appropriate amount of additional sterile degassed water. Cross-linking of the hydrogel is achieved through thiol-based chemistry, a reaction which takes only 10-20 minutes and occurs at physiological pH. Engelbreth-Holm-Swarm (EHS) growth-factor-reduced extracellular matrix (Sigma-Aldrich) was used in its native composition. The EHS matrix was placed in liquid state into the PDMS wells at 4°C and allowed to gel at incubation temperature.

For conditions containing hydrogel (both HA and EHS), cells were suspended in the hydrogel and the hydrogel was carefully placed into the PDMS well, up to but not exceeding the height of the PDMS rings. Diffusion modeling demonstrated sufficient capability for both oxygen and nutrients to diffuse through the hydrogel construct. For 3D conditions with electrospun fibers, the fibers were embedded through the middle of the hydrogel scaffold by carefully placing the hydrogel around the fiber layers within the PDMS construct. Slow micropipette extrusion of the hydrogel around the fibers enabled placement with minimal disruption of the fibers. The hydrogel was allowed to set for 5-10 minutes before loading feeding media around the hydrogel construct in order to allow the hydrogel to set around the fibers without destroying them and without washing away hydrogel in the new media.

HA hydrogels were first mixed without cross-linker in aliquots for each set of conditions, and each aliquot of HA or EHS hydrogel was mixed with the appropriate number of cells, as calculated from the hemocytometer reading of the number of cells/ml in the cell suspension solution. For 2D conditions, a calculated amount of the cell suspension solution (containing sufficient cells for all 2D conditions) was centrifuged at 300g for 3 minutes, after which the supernatant was removed and the cells were re-suspended in media at approximately 400 cells/µl, and 35 µl of this media was deposited within each cylindrical PDMS well onto 2D coverslips. For 3D hydrogel conditions, calculated amounts of the cell solution (containing sufficient cells for 3D EHS conditions and for 3D HA conditions) were each centrifuged separately at 300g for 3 minutes, after which the supernatant was removed and the cells were carefully re-suspended in either the EHS hydrogel or the HA hydrogel. The cells and cross-linker were carefully mixed with the HA hydrogel before placement in the PDMS constructs, and the amount of cross-linker solution used was the same amount as would be required to produce 1% HA hydrogel (0.25 ml cross-linker per 1 ml HA solution) but with the total hydrogel diluted to 0.75%. For every hydrogel condition, 20,000 cells in 50 µl of hydrogel was placed in each well.

After seeding both the cellularized hydrogels and the cells in media, cells were cultured in complete media with supplements of 10µM all-trans-retinoic acid, 50ng/ml BDNF, 10ng/ml NGF, and 2% B27, and cells were fed every two days for a period of 4 days. Cell constructs were then carefully washed in PBS and fixed in 4% paraformaldehyde for 20 minutes at room temperature, then washed three times with PBS. Experiments were carried out twice with three wells per condition.

*Imaging*

Imaging of samples was performed on a Nikon TiE2000 inverted fluorescent microscope with NIS Elements software. In order to image the cells with and without the fibers, images were captured in a green fluorescence channel (FITC 470 nm) for visualization of eGFP in the cells and in an optical channel for imaging fibers. Fibers did not exhibit autofluorescence (see Figures 3-5), which enabled analysis of colocalization of pixel intensities between the two channels at high resolution and careful marking of neuronal morphology at high magnification. Images were taken in the focal plane of the fibers and Z-stack images were taken of 3D conditions, and images were adjusted for optimal recognition of cell borders or neurite morphology using look-up tables (LUTs). Image parameters were calibrated to know the number of micrometers per pixel, which enabled detailed measurements of neurite lengths with the software, and the images were measured and analyzed at the cellular and sub-cellular levels.

*Morphologic Analysis*

Morphological characteristics were analyzed with NIS Elements software, with data including neurite lengths, neurite numbers, neurite angles, neurite branching, and direct neurite/fiber association length. Neurites were further classified as primary (originating from the cell body), secondary (originating from another neurite), and longest neurite (the single longest continuous neurite from each cell body). Morphological labeling was achieved by viewing and marking channel features in sequence, first viewing the neuronal channel alone (displayed in green) and marking cell borders and neurites in red lines; the cellular channel was then closed and the fiber channel was opened (displayed in blue), allowing analysis of colocalization, measured as overlap between nanofibers and neurite morphologies, and these colocalizations were marked with white lines, thereby representing where neurites were directly tracking fibers. The software automatically recorded lengths and angles of all line segments, and classifications of neurite type and neurite features were documented manually in the same data spreadsheets.

For cells to be measured and analyzed, they had to be in the focal plane of the fibers and have identifiable cellular borders such that any neurites could be measured from the cell. No nuclear stain was used since cells were found to have identifiable nuclear densities from eGFP expression and cells could be reliably delineated by cellular borders, and this avoided problems of optical interference inherent in 3D hydrogel cultures with overlying layers of cells while also keeping open color channels needed for marking morphology. In conditions with nanofibers, only cells that were in contact with the fibers, or cells that made contact with a fiber via a neurite, were counted, since cells that were not touching the fibers could be expected to behave similarly to the paired hydrogel conditions without fibers. One hundred or more cells were characterized from each condition ($n \geq 100$), sampled from each of the triplicate repeats of the condition sets. In order to perform t-tests, the lengths and angles of neurites, as well as the tracking affinity factor (described below) were calculated per cell, enabling sample data to be analyzed for mean, standard deviation, and significance. All results were compared using ANOVA and then two-tailed t-tests assuming unequal variances to examine whether distribution characteristics of the aforementioned variables differed significantly between each condition.

It is important to use methods of neurite analysis that allow meaningful interpretation. Multiple methods were used to accurately describe fiber-tracking behavior by neurites. In conditions with fibers, the angle of each neurite segment was measured with reference to the fibers with which it was in contact or to the nearest fibers, with the fiber being defined as 0°, and in conditions without fibers, 0° was defined to the right of the image as on Cartesian coordinates. Angle measurements were constrained to be between 0° and 90°, meaning that the angle was measured as the smallest angle to the fiber or to the 0° axis, as this method provides a meaningful description of whether the neurites were tracking the fibers (e.g., if a neuron had two neurites extending at 0° and 180°, then the data would show the average neurite angle to be at 90° relative to the fiber, which is misleading; instead, constraining the angle between 0° and 90° shows the average angle to be 0°, which accurately depicts the neurites' orientation relative to the fibers). An average angle of 0° thus represents alignment of neurites and fibers, while an average angle of 90° indicates that neurites and fibers are perpendicular. For graphing neurite angle distributions, neurite angles were grouped into 5° increments, ranging from 0-90°. Angles between 0 to 0.1° were classified as exactly 0° in order to demonstrate the group of neurites that were exactly aligned with fibers. Thus the first group at 0° represents only 0-0.1° (direct alignment), the second group represents 0.1-5°, and every other group represents a 5° span.

In conditions with fibers, the "Tracking Affinity Factor" (TAF) was measured as the distance over which each neurite traveled in direct contact with the fiber (recorded for each sampled cell) compared to total neurite length (also recorded for each sampled cell), thereby providing an accurate description of how much the neurites tracked the fibers. A tracking affinity factor of 100% thus indicates that 100% of the neurite was in direct contact with the fiber, whereas 0% would indicate that none of the neurite lengths of a cell followed a fiber. The TAFs of each sampled cell were then averaged and compared for each condition. This factor is not applicable to control conditions without fibers.

$$Tracking\ Affinity\ Factor\ (\%) = \frac{Length\ of\ Avg\ Neurite\ Colocalization\ with\ Fiber\ (Per\ Cell)}{Average\ Total\ Length\ of\ Neurites\ (Per\ Cell)}$$

## Results and Analysis

### Analysis of Electrospun Fibers

Fiber diameters ranged from approximately 600-1200 nm, with the average diameter being 841 nm for plain PCL fibers (Table 1 and Figure 2). Laminin coating of PCL fibers resulted in a small but significant increase of 228 nm in fiber diameter. No notable surface features were observed on any of the fiber types.

| Fiber width (nm) | | Comparisons (p-value) | |
|---|---|---|---|
| Plain PCL polymer: | 841 ± 112 | PCL vs PCL/Gelatin | 0.05 |
| PCL/Gelatin Mixture: | 782 ± 131 | Laminin vs PCL/Gelatin | <0.01 |
| Laminin-Coated PCL: | 1069 ± 126 | Laminin vs PCL | <0.01 |

Table 1: Mean fiber widths ± standard deviation (in nm), with statistical comparisons (p-values).

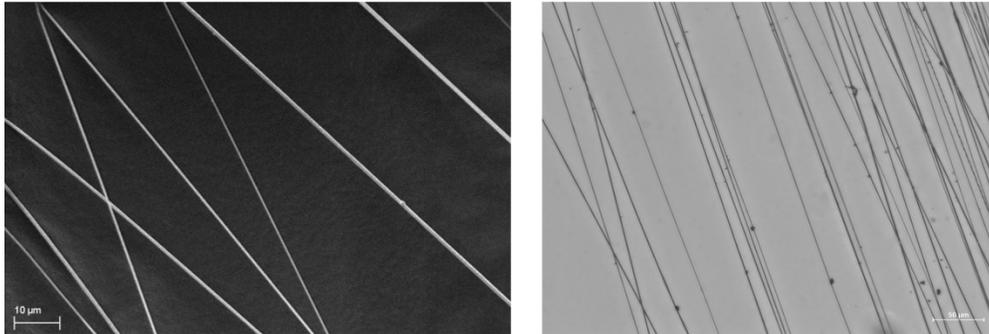

Figure 2 (a-b): Scanning electron microscope image of electrospun fibers with laminin coating (a), and optical microscope image of nanofibers suspended in HA hydrogel (b).

### Analysis of Neurite Direction

A broad view of a 3D cellularized construct with fibers is shown at low magnification in Figure 3. Examples of the morphological analysis of the 3D HA hydrogel with laminin-coated nanofiber scaffolding are shown at high magnification in Figures 4 and 5. Neuronal morphologies on the control condition of 2D laminin surfaces are also shown in Figure 6. For conditions where fibers were present, neurite angles tended towards 0°, meaning that the neurites have a strong tendency to orient in the direction of aligned fibers, and this tendency held true for all fiber compositions in both two and three dimensions (Figure 7). The large distribution of neurite angles within 0-5° also suggests that even when neurites do not directly align with fibers, they still tend to run in the same direction, such as when travelling between fibers.

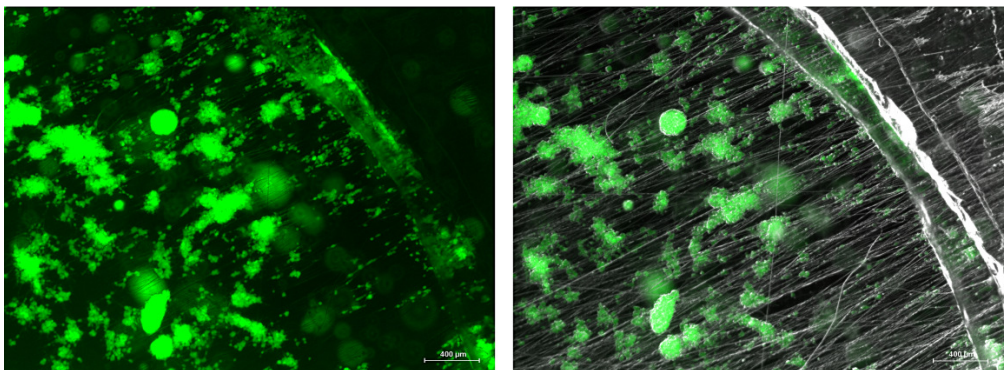

Figure 3 (a-b): Overview of the composite 3D hydrogel nanofiber construct (40X magnification) showing the tendency of neuronal cells to cluster in globules while suspended at varying levels within the hydrogel. However, at the level of the laminin-coated nanofibers running through the hydrogel, separation and extension of the small individual neurons can be seen along the fibers. The edge of the construct can be seen at the upper right. (a) Cells alone are seen in the eGFP fluorescence channel, and (b) the merging of optical and fluorescent channels allows fibers and cells to be seen together.

The presentation of raw angles of all neurites provides an intuitive view of the alignment behavior, but does not account for the length of each neurite, its adherence to fibers, or its overall contribution to neurite tracking. As an example, the neurite angle distribution appeared similar for many conditions, but the amount of direct tracking of fibers by neurites was significantly different between conditions, as measured by the tracking affinity factor. Nevertheless, these data show that laminin-functionalized fibers in HA hydrogel provide the greatest degree of neurite alignment with fibers, with 29% of neurite segment angles directly matching the direction of fibers and 67.4% within 5 degrees of the fiber direction (p<0.001 compared to all other conditions).

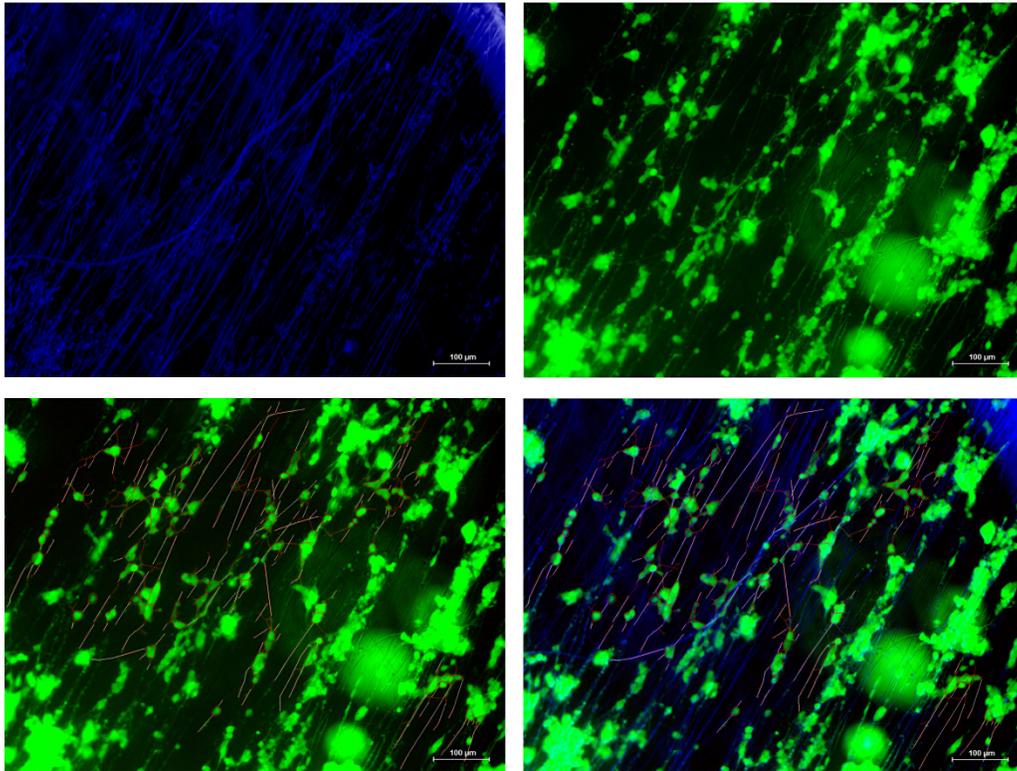

Figure 4 (a-d): Images of neuronal cultures in 3D HA hydrogel with laminin-coated nanofibers (100X magnification). The channels of cells and fibers are separated, including images of fibers alone (a) and eGFP-expressing cells alone (b), and then with morphological labeling of cells in (c) and labeling of neurite tracking of fibers in (d). Blue shows the laminin-coated nanofibers, red marks neurites and cell borders, and white marks nanofibers in contact with neurites. Unfocused cell clusters in the hydrogel out of the focal plane can be seen at the bottom right and left corners of the image.

*Tracking Affinity Factor*

All conditions with fibers had some amount of direct fiber tracking by neurites, as shown in Figure 8. Similar to the measure of neurite angles, the tracking affinity factor was highest for the condition of laminin-coated fibers in HA hydrogel. In this condition, cells produced an average of only 3.4 neurites per cell, but the average length of each neurite was 31.1 microns, of which an average of 66.4% directly tracked fibers (66.4% vs. 21.2% on 2D laminin-coated fiber surfaces, p<0.001). Figure 8 also shows that laminin functionalization of fibers on 2D laminin-coated surfaces produced less direct neurite tracking than plain fibers on the same surface (21.2% vs. 39.2% on plain fibers, p=0.007). However, laminin functionalization significantly increased neurite tracking along nanofibers in 3D HA constructs, with 66.4% of neurite lengths directly tracking the laminin-coated fibers, which was a 65.2% relative increase in neurite tracking compared to plain PCL fibers in the same 3D HA constructs (66.4% vs. 40.2%, p<0.001) and a 213.2% relative increase over the same fibers on 2D laminin-coated surfaces (66.4% vs. 21.2%, p<0.001).

*Neurite Length and Number*

Laminin-coated nanofibers in 3D HA constructs significantly increased the average length of neurites compared to both 2D and 3D control conditions without fibers, as shown in Figure 9 (31.1 vs. 19.5 μm on the 2D laminin-coated surface without fibers, p<0.001, and 31.1 vs. 2.7 μm in the HA hydrogel without fibers, p<0.001). Laminin functionalization of fibers in 3D HA constructs doubled average neurite length over plain PCL fibers in the same constructs (31.1 vs. 15.5 μm, p=0.021) and also increased average neurite length by 1051.9% compared to the same constructs without fibers (31.1 vs. 2.7 μm, p<0.001).

Although laminin-functionalization greatly enhanced neurite outgrowth in 3D HA constructs, neurite outgrowth also occurred in the presence of plain PCL fibers, which showed a tremendous effect in extending average neurite length by 473.4% compared to the same HA constructs without fibers (15.5 vs. 2.7 $\mu$m, p<0.001). Gelatin functionalized nanofibers appeared somewhat inhibitory to neurite outgrowth, showing shorter average neurite lengths than plain fibers in both EHS and HA hydrogel (12.8 vs. 15.5 $\mu$m, p=0.045 in HA, and 18.8 vs. 22.7 $\mu$m, p=0.028 in EHS).

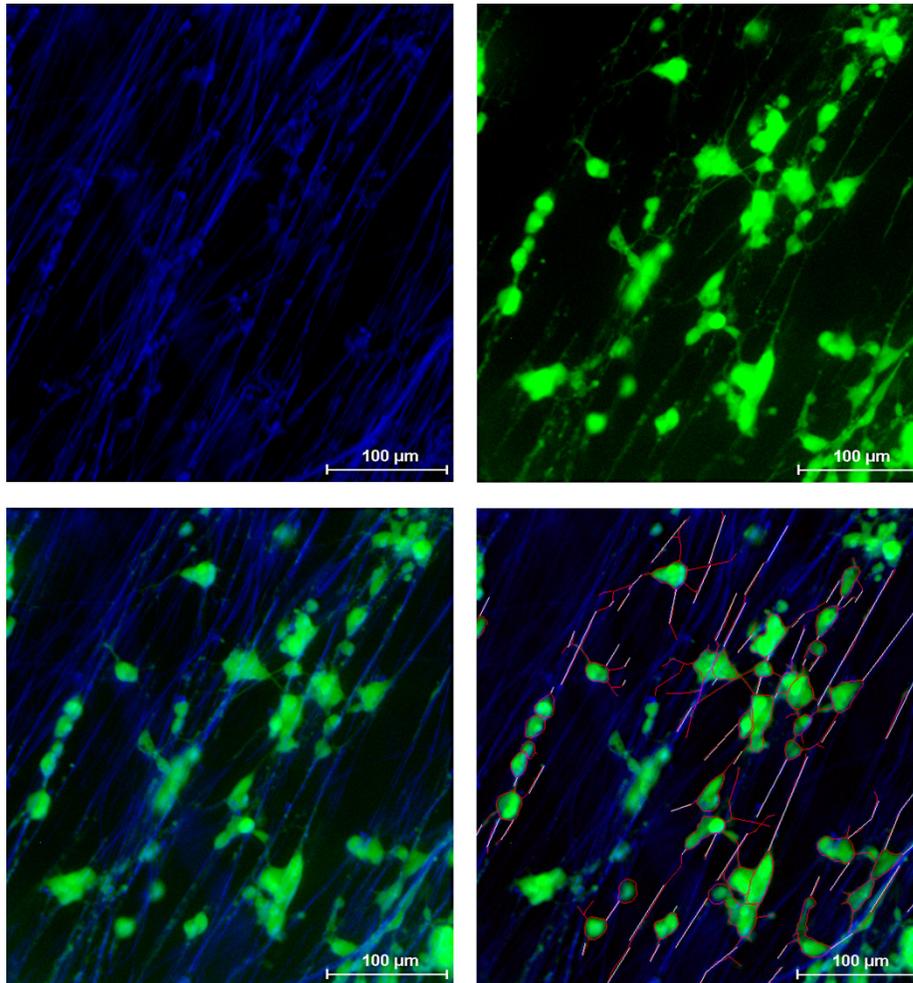

Figure 5 (a-d): Magnified views of the same field shown in Figure 4, demonstrating neurite extension and overlap along the fibers. Images may best be viewed by sequentially flipping through the images since the overlapped images can obscure neurites that are about the same thickness as the fibers.

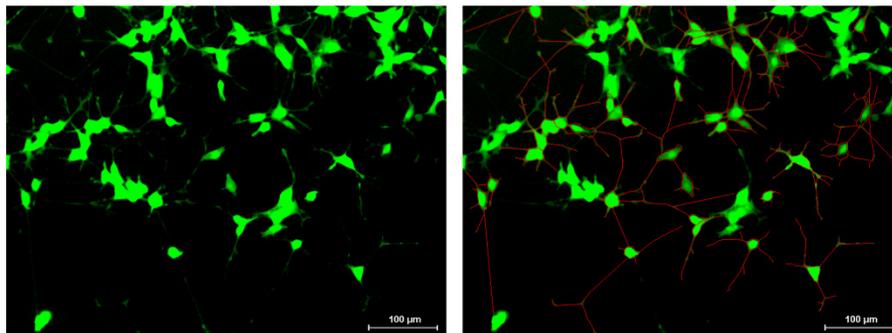

Figure 6 (a-b): Examples of neuronal cultures on a 2D laminin-coated surface (a), and with labeled morphology of neurites and cell borders in red (b).

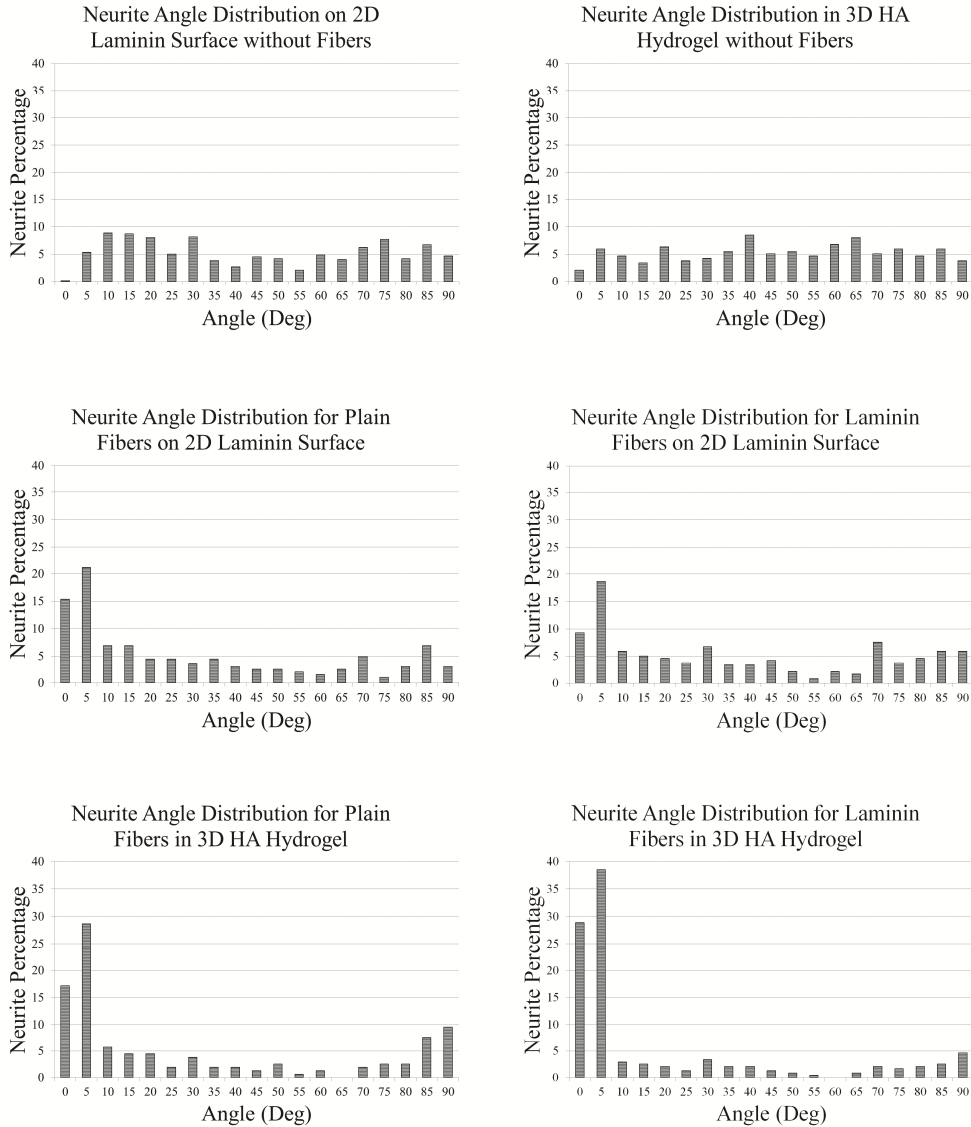

Figure 7 (a-f): Results in Neurite Angle Distributions. Top row– Control conditions, showing random distribution of neurite angles in 2D and 3D conditions without fibers. Middle row—2D conditions with plain PCL fibers or laminin-coated fibers. Bottom row– 3D nanofiber constructs, showing significant alignment of neurites with nanofibers, as confirmed by $\chi^2$ testing: p<0.001 for (e) vs. (a), (e) vs. (b), (f) vs. (a), (f) vs. (b), (f) vs. (d), (d) vs. (b), (c) vs. (a), and (e) vs. (f); p=0.035 for (e) vs. (c). There was no significant difference between (a) vs. (b) or (c) vs. (d).

Neurite extension was also influenced by the use of HA hydrogel as shown by the fact that 3D HA constructs with laminin-coated fibers significantly increased the average length of neurites by 65.6% over the same laminin fibers on 2D laminin surfaces (31.1 vs. 18.8 µm, p<0.001) and by 59.3% compared to 2D laminin-coated surface without fibers (31.1 vs. 19.5 µm, p<0.001). The 3D EHS hydrogels, however, did not show significant differences in neurite length compared to 2D conditions with the same fibers. The use of HA hydrogel increased neurite length by 117.5% over EHS hydrogel with the same laminin-coated fibers (31.1 vs. 14.3 µm, p<0.001) but also resulted in fewer neurites per cell (3.4 vs. 7.1, p<0.001).

When looking at just the longest neurite per cell in each condition, the effect of laminin-coated nanofibers becomes even more apparent (Figure 10). Laminin functionalization of fibers in 3D HA constructs increased the average length of the longest neurites by 105.9% over plain PCL fibers in the same 3D HA constructs (56.2 vs. 27.3 µm, p<0.001), by 36.1% over 2D laminin surfaces with laminin-coated fibers (56.2 vs. 41.3 μm, p=0.002), and by 76.2% over 2D laminin surfaces without fibers (56.2 vs. 31.9 μm, p<0.001).

In summary, it was found that laminin-coated nanofibers in 3D HA hydrogels enabled significant alignment of neurites with fibers (Fig. 7), resulted in significant neurite tracking of nanofibers (Fig. 8), and significantly increased the distance over which neurites could extend (Figs. 9 and 10).

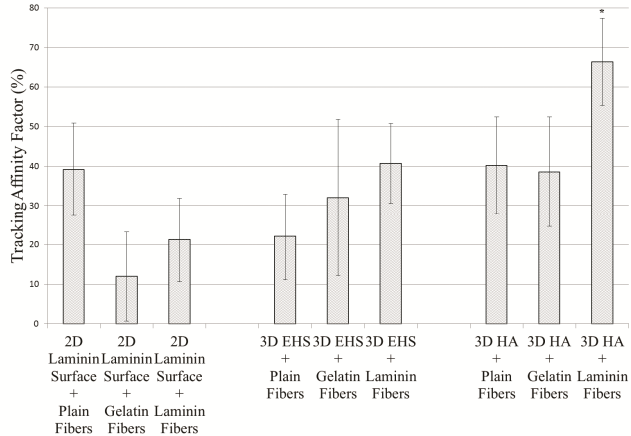

Figure 8: Tracking Affinity Factor (TAF) showing the percentage of neurite length directly in contact with a fiber for all conditions with fibers. 2D surfaces are shown in the left group, 3D EHS hydrogel conditions in the middle group, and 3D HA hydrogel conditions in the right group. In HA hydrogel, neurites on laminin-coated fibers exhibited a TAF increase of 26.2 (or 65.2% more relative tracking) compared to plain PCL fibers (*p<0.01 for '3D HA + Laminin Fibers' compared to all 2D conditions and compared to all plain fiber conditions). Differences between each laminin-coated fiber condition were also significant (p=0.002 for '2D + Laminin Fibers' versus '3D EHS + Laminin Fibers' and p<0.001 for '3D EHS + Laminin Fibers' versus '3D HA + Laminin Fibers'). Error bars = ±S.D.

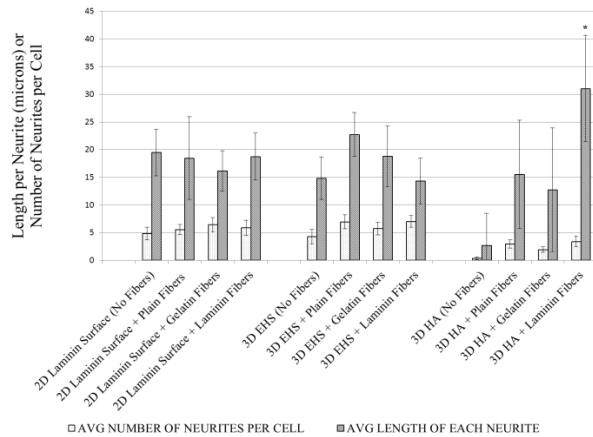

Figure 9: Average length of all neurites and the average number of neurites per cell are shown. Left grouping: 2D conditions, which exhibited similar lengths and numbers of neurites. Middle grouping: EHS hydrogel conditions. Right grouping: HA hydrogel with fibers, which exhibited the greatest change in neurite lengths depending on the presence and type of nanofiber scaffold. Error bars = ±S.D.

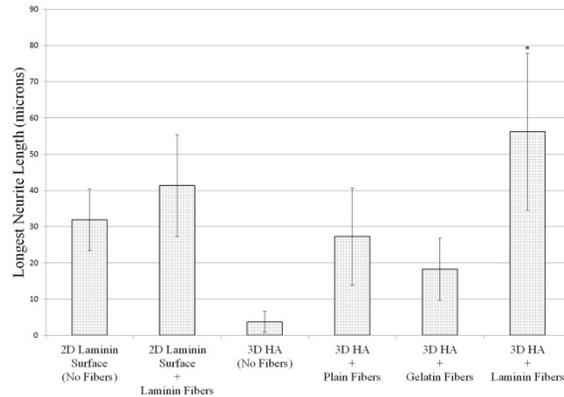

Figure 10: Comparison of the average length of the longest neurite per cell: The use of laminin-coated fibers in 3D culture increased longest neurite length per cell compared to the 2D control without fibers by 76.2% (*p=0.002 when comparing the '3D HA + Laminin Fibers' condition to the '2D Laminin Surface + Laminin Fibers' condition and p<0.001 when comparing to that of every other condition). Error bars = ±S.D.

## Discussion

This work demonstrated the ability to create patterned 3-dimensional neuronal constructs using a combined system of hydrogel and electrospun fiber scaffolding. It was found that aligned nanofibers enabled significant directional control of neurite outgrowth both on 2D surfaces and in 3D mediums, and furthermore, that the use of laminin-functionalized nanofibers in 3D HA hydrogel significantly enhanced neurite length and neurite tracking along the nanofibers. Low neurite outgrowth was expected in the 3D HA hydrogels without fibers since hyaluronan hydrogels do not possess functional groups to which cells can easily attach, but the addition of nanofibers through a hydrogel with weak cell attachment points significantly enhanced neurite outgrowth along the fiber.

In these experiments, all conditions with nanofibers, regardless of fiber functionalization, exhibited some amount of direct fiber tracking by neurites and angle orientation of neurites along the fibers, demonstrating that single polymer aligned nanofibers alone can create favorable cues for neurite tracking. The molecular mechanisms of mechanical signaling phenomena, particularly for neuronal cells, are not well understood, but it appears that some aspects of cell morphology and neurite extension are based on structural cues in addition to biological cues. The enhancement of this effect by coating fibers with the biologic adhesion molecule of laminin shows that neurite tracking can be further improved by combining structural and biochemical cues into functionalized fibers. This ability of fibers to direct neurite extension may be related to the phenomenon observed in early neurodevelopment where cells and neurites extend along radial glial fibers that are of a similar scale as the fibers used in this study [67-70].

It was somewhat puzzling, however, that in comparing neurite tracking of plain fibers versus laminin-coated fibers on 2D laminin surfaces, neurite tracking was significantly lower on laminin-coated fibers in the 2D conditions (21.2% vs. 39.2% on plain fibers). Both 2D conditions showed neurite angle distributions and neurite lengths that were nearly identical and not significantly different from each other (Figure 7 and Figure 9), suggesting that both types of fibers are equally capable of orienting neurites to their direction. Additionally, the fact that these 2D conditions demonstrated TAF values of 20-40% indicates that significant tracking of fibers is occurring, in accordance with previous reports [31,56,62]. Importantly, because the TAF represents a percentage of neurite length that is in direct contact with a fiber, it should also be interpreted in the context of total neurite length and number, since it could be deceptively lowered by an increase in total neurite lengths rather than by a decrease in direct neurite tracking of fibers. Nevertheless, in this case total neurite lengths among all 2D conditions were not significantly different, with averages among the 2D conditions ranging from 95 μm (for no fibers) to 110 μm (for laminin fibers), meaning that the difference in TAF between plain fibers and laminin fibers on a 2D surface was in fact due primarily to increased neurite tracking along plain fibers. The reason for this appears to be that neurites on the laminin-coated fibers with laminin-coated surfaces can more easily cross over and deviate from fibers since all surfaces have the same coating, whereas neurites encountering the differential surface of an uncoated polymer fiber might instead track along the fiber for some distance rather than cross over it. This phenomenon may also be more evident when aligned fibers are more sparsely spaced on 2D surfaces, as they were in these experiments, since cells may contact dual surfaces of smooth glass and intermittent fibers each with the same biologic surface coating, whereas denser fiber bundles might exert more topological effects and mechanical constraints that help the alignment of neurites and prevent wandering across flat surfaces. The spacing of topographical features has been shown to affect many types of cell growth and migration [71], but the role that spacing of ridge-like fibers on a smooth surface plays in directing neurite outgrowth is not yet fully understood. It is also not known whether the small but statistically significant difference in diameters between fiber types could have independently influenced the results—the small difference and overlapping distribution of fiber diameters between conditions makes this unlikely, but further study is still needed to better understand how fine details of controlled nanotopology can affect neurite outgrowth [41-42,45,47-48,51,57,61].

Similar to the above, in EHS hydrogels it appears that the tracking affinity factor is not particularly high even for laminin-coated fibers. This could have been due to the fact that EHS hydrogels enabled higher numbers of neurites per cell compared with the HA hydrogel (7.1 versus 3.4), meaning that the total neurite length might have been higher and the tracking affinity factor would therefore be lower; however, comparison of total neurite lengths per cell showed that lengths were essentially the same, with an average of 101 μm in the EHS hydrogel and 105 μm in the HA hydrogel. Therefore in this case the tracking affinity factor is an accurate representation of the ability of neurites to track the laminin-functionalized fibers in the different hydrogels, where the average neurite tracking length per cell was 63.2% higher in the HA hydrogel than in the EHS hydrogel for laminin-coated fibers (TAF 66.4 vs. 40.7, p<0.001). The lower tracking of functionalized fibers in EHS with the same average total neurite length, along with the higher neurite length in EHS without fibers compared to HA without fibers, further suggests that the EHS matrix contains abundant molecular attachment points by which the cells may adhere and extend neurites, thereby enabling the neurite outgrowth to occur without necessarily following the fiber scaffold. These results together suggest that the best method for controlling neurite direction and enhancing neurite length is to use functionalized nanofibers within a hydrogel that is relatively inert in cell attachment characteristics.

This work also provides new approaches for controlling electrospun fiber alignment and for analyzing neurite directivity and alignment of neurites on external cues. With the advent of increasingly complex approaches to engineering neural tissue and drug delivery to neural tissues, the need for meaningful analysis of neuronal morphology will increase over the coming years, and the methods presented here present a useful approach for reporting and interpreting morphological results.

An effective approach in regenerating functional neural tissue will require a combination of cellular, molecular, and structural cues to guide the regeneration of functional neural tissue. The combination of electrospun fiber scaffolding with cellularized hydrogels in this study demonstrates integration of structural, cellular, and biochemical signals into a single construct, and these constructs also hold several potential benefits for implantable neural tissue grafts, such as being biocompatible, scalable in complexity, protective to neuroglial cell cultures, and gel-stabilized for direct implantation. This also provides evidence that further fiber functionalization approaches may be implemented in more complex patterning and opens opportunities to investigate the application of axonal guidance signals in 3D constructs. This approach may also be viable for creation of functional neural networks with patterned neural circuitry and the ability to direct neurodevelopment of stem cells [28-29,37,48-50].

The creation of functional neural tissue constructs, however, will still require overcoming many challenges in order to prepare the technology for successful clinical implementation. Neural regeneration is notoriously difficult for a variety of reasons, particularly due to the complexity of the neural environment and the cellular and subcellular architecture. In addition, diffusion limitations may dictate size constraints of the constructs and may require additional features, such as tiered stacking of constructs with fluid vents or channels between layers, in order to help circumvent diffusion limitations. Further problems may arise with integration of neural connections between the implanted construct and the neural tissue itself, including problems with coaptation of axonal pathways or deleterious side effects of allodynia, hyperaesthesia, or hypertonia seen in some spinal cord regeneration efforts [72-73]. Nevertheless, the use of patterned scaffolding matrices may help overcome these problems by guiding neurite extension, targeting proper neural connections, and providing a local environment that supports cell survival, proliferation, and differentiation. In fact, recent efforts using the combination of neural stem cells within unpatterned fibrin scaffolds containing growth factors has demonstrated remarkable ability to extend functional axonal connections across lesion sites in spinal cord injuries in rats [74], and cross-linked hyaluronic acid hydrogels have shown improved survival of stem cells implanted in the brain [75-77], further suggesting that combinations of cells, scaffolds, and hydrogels may indeed be an ideal approach for regeneration and reconstruction of neural tissue.

Altogether this work demonstrates a novel approach for creating 3D patterned scaffolds for guiding neurite outgrowth of neuronal cultures. 3D constructs hold important potential for implantation as neural tissue grafts, and the integration of nanofiber scaffolding holds important potential for replicating specific neuroanatomical pathways and guiding neural network formation down to the subcellular level. This study serves as a proof-of-concept that 3D tissue architecture can be assembled with seeding of induced stem cells on functionalized scaffolding. Future work will attempt more complex patterning to replicate neuroanatomical structures of the spinal cord, cortex, hippocampus, and other structures, as well as seeding with induced pluripotent stem cells, thereby opening the possibility of treating neurological tissue damage with a patient's own cells on scaffolding that replicates innate neural architecture. These scaffolds would be directly applicable to specific areas of damage that are amenable to graft implantation, particularly spinal cord injuries, nerve injuries, tumor resection sites, and areas of cortical damage. This work may also play an important role as a novel therapeutic approach to many diseases of neural tissue, including stroke, traumatic brain injury, and neurodegenerative diseases.

## Appendix: Modeling Fiber Alignment

A formula was derived to estimate the appropriate collection drum rotation rate that would best produce aligned fibers. The extrusion rate of polymer solution is set at the syringe pump, producing a narrow stream of fluid exiting out the nozzle, and this stream can be assumed to be cylindrical in nature. The velocity of this stream should be less than or equal to the surface velocity of the rotating collection drum in order for the fibers to remain straight, otherwise the fibers may begin to wander or collect unevenly. The thickness of the final dried fibers can be measured, and it may be assumed that the entirety of the solvent evaporates. If it is assumed that the composition of the final fibers is condensed and of homogeneous density, and if it is assumed that the volume (and therefore the diameter) of the stream is related to the volume (and therefore diameter) of the fibers plus the volume of solvent that was evaporated, then the diameter of the stream ($D$, in microns) can be estimated from the diameter of the dried fibers ($d$, in microns) by the formula

$$D = \frac{d}{c} \quad (1)$$

where $c$ is the concentration of polymer in the solvent (e.g., a 6% solution means $c$=0.06). Let $G$ equal the diameter of the collecting drum (in cm), and let $R$ equal the extrusion rate (volume per time, measured in ml/min). The average velocity of the stream can then be found by simply equating the volume of solution ($V$) extruded per unit time ($t$) with the volume of the stream collected on the drum, and the volume of the stream can be equated to a cylinder of diameter $D$ and height $L$.

$$R = \frac{V}{t} = \frac{\pi \left(\frac{D}{2}\right)^2 L}{t} \quad (2)$$

The value of $R$ is simply set on the extrusion syringe pump. The volume extruded may then be divided by the cross-sectional area of the fiber $\left(\frac{\pi D^2}{4}\right)$ in order to find the length of the fiber, $L$, that must be extruded per unit time, which also equates to the average velocity of the stream ($U$):

$$U = \frac{L}{t} = \frac{4R}{\pi D^2} \quad (3)$$

The minimum rotational rate (in rotations per minute, RPM) for collecting aligned fibers can then be found by dividing the jet stream velocity ($U$) by the circumference of the collecting drum ($\pi G$) and inserting a correctional factor for the units ($10^8 \frac{\mu m^2}{cm^2}$).

$$Minimum\ Rotational\ Rate\ = \frac{U}{\pi G} \cdot (10^8) \quad (4)$$

Finally, substituting (3) into (4) provides the following convenient formula that matches the velocity of the jet to the linear surface speed of the rotating drum, thereby calculating the minimal rotational rate needed to maintain straight fibers, where $R$ (in ml/min) and $G$ (in cm) are set by the experimenter, and $D$ (in microns) is found empirically per above:

$$Minimum\ Rotational\ Rate\ = \frac{4R}{\pi^2 D^2 G} \cdot (10^8) \quad (5)$$

Thus formula (5) provides an estimate of the appropriate rotational rate that is normalized to the drum diameter, and the rotational rate is set by including a rotational counter in the equipment setup. In these experiments, the minimal rotational rate was calculated to be 500 rpm, and although adequate aligned fibers were obtained at this speed, optimal results were obtained with the rotational rate at 1000 rpm. The variation is most likely due to the fact that the fiber is not perfectly dense and homogeneous, making a width measurement that is larger than that provided under the assumptions above, thus underestimating the minimum rotational rate. In addition, a rotational rate faster than the minimum rotational rate exerts a tensile load on the fiber, adding the benefit of stretching the fiber along the surface, but if the rotational rate is too fast, fibers may be broken or may detach from the collecting surface due to rotational and shear forces.

## Acknowledgments

I would like to express gratitude and appreciation to many, including Dr. Julian George, Dr. Hua (Cathy) Ye, and Prof. Zhanfeng Cui at the University of Oxford Institute of Biomedical Engineering for their kind support, to Dr. David Nagel, Dr. Eric Hill, and Prof. Michael Coleman at Aston University Research Centre for Healthy Ageing for the cell line, and to Dr. Kalin Dragnevski at the University of Oxford Department of Engineering Science for the use of the scanning electron microscope.

## Author Disclosure Statement

No grant support was provided for this work. The author declares that he has no competing interests.